\documentclass[12pt]{iopart}
\usepackage{iopams}  
\usepackage{setstack}
\usepackage{graphicx}
\usepackage{float}

\begin{document}

\title[Hidden degrees of freedom and fluctuation theorems: an analytically solvable model]{Hidden slow degrees of freedom and fluctuation theorems: an analytically solvable model}

\author{Marcel Kahlen and Jannik Ehrich}

\address{Universit\"at Oldenburg, Institut f\"ur Physik, 26111 Oldenburg, Germany}
\ead{marcel.sebastian.kahlen@uni-oldenburg.de, jannik.ehrich@uni-oldenburg.de}
\vspace{5mm}
Version: \today

\begin{abstract}
In some situations in stochastic thermodynamics not all relevant slow degrees of freedom are accessible. Consequently, one adopts an effective description involving only the visible degrees of freedom. This gives rise to an apparent entropy production that violates standard fluctuation theorems. We present an analytically solvable model illustrating how the fluctuation theorems are modified. Furthermore, we define an alternative to the apparent entropy production: the marginal entropy production which fulfills the fluctuation theorems in the usual form. We show that the non-Markovianity of the visible process is responsible for the deviations in the fluctuation theorems.
\end{abstract}

%
%
%
\maketitle
%
%

\section{Introduction}
Stochastic Thermodynamics allow the study of small-scale systems driven far away from thermal equilibrium~\cite{Jarzynski2011, Seifert2012}. This is usually achieved by describing the properties of interest as stochastic processes. In this context fluctuation theorems play a central role, as they allow to link quantities obtained from nonequilibrium transformations to equilibrium system properties~\cite{Jarzynski1997,Crooks1999}. The formalism of stochastic thermodynamics commonly assumes a time-scale separation between the slow observed degrees of freedom and the fast unobserved variables which are assumed to be equilibrated~\cite{Esposito2012}.

However, this assumption cannot always be fulfilled. For example, in molecular folding-unfolding experiments in which multiple laser traps are being used, it is not always practical or possible to observe the dynamics in both traps thus rendering a degree of freedom hidden from the observer~\cite{Ribezzi-Crivellari2014, Alemany2015}. Moreover, studies on molecular motors often rely on the attachment of beads to the system under study, which results in a joint stochastic system of motor and bead~\cite{Kolomeisky2013}. Since only the bead is observed, the degrees of freedom comprising the motor are hidden from the experimenter. It is therefore important to study the influence of hidden degrees of freedom in the context of stochastic thermodynamics.

A very general approach to this problem relies on the interpretation of any deviations of measurable quantities as measurement errors and studying their impact on fluctuation relations~\cite{Garcia-Garcia2016,Waechtler2016}.

Further, one may obtain an effective description of the visible degrees of freedom by employing a coarse-graining scheme~\cite{Esposito2012, Bo2014} which lumps together several (hidden) microstates into few (observable) mesostates. An early theoretical study on the impact of coarse-graining on fluctuation relations has been carried out by Rahav and Jarzynski~\cite{Rahav2007}. An experiment of two magnetically coupled colloidal particles of which one is hidden from the observer has been realized by Mehl \emph{et al}~\cite{Mehl2012} and was recently further analyzed theoretically by Uhl \emph{et al}~\cite{Uhl2018}. By employing coarse-graining the authors define an \emph{apparent entropy production} for the resulting effective process. When evaluated, the fluctuation theorems for this quantity deviate from the usual form expected for the effective process. In the same spirit Chiang \emph{et al}~\cite{Chiang2016} have experimentally investigated the fluctuations of entropy production in a driven RC-circuit coupled to another hidden circuit with similar results.

The question of how to appropriately split the entropy production for systems with interacting degrees of freedom has attracted some attention recently~\cite{Hartich2014, Horowitz2014, Ehrich2017} and there are also alternative definitions of coarse-graining applicable to networks of states of discrete Markov processes~\cite{Puglisi2010, Altaner2012, Esposito2015}. Furthermore, there have been efforts to formulate an effective thermodynamic description for these systems if not all transitions are observed~\cite{Shiraishi2015,Polettini2017,Bisker2017}.

However, what is still lacking is a sufficiently simple and thus analytically tractable model system which illustrates the effect of coarse-graining in a system with hidden slow degrees of freedom. Such a model has the added benefit of being able to pinpoint why the apparent entropy production violates standard fluctuation theorems.

The aim of this paper is to 1) present such a model system with a hidden degree of freedom, obtain an effective description of the observed degree of freedom and analytically calculate the fluctuation theorems for the apparent entropy production; 2) offer a complementary \emph{marginal entropy production} which fulfills the fluctuation theorems in their standard form; and 3) identify the difference between these two methods of effective description.

\section{Fluctuation relations and coarse-graining}
Let a system with two degrees of freedom be described by a bivariate Markov process $\{x(t),y(t)\}$. The evolution of the joint probability of the entire process shall be given by a Master (or Fokker-Planck-) equation:
\begin{equation}\label{eqn_jointMasterEquation}
\partial_t\, p(x,y;t) = \mathcal{L}(t)\,p(x,y;t),
\end{equation}
where $\mathcal{L}(t)$ is the generator.

We now consider trajectories of length $T$ of the joint process and define a stochastic entropy production $\sigma[x(\cdot),y(\cdot)]$~\cite{Seifert2005}:
\begin{equation}\label{eqn_defJointEP}
\sigma[x(\cdot),y(\cdot)] := \ln\frac{p[x(\cdot),y(\cdot)]}{\bar{p}[\bar x(\cdot),\bar y(\cdot)]},
\end{equation}
where $p[x(\cdot),y(\cdot)]$ is the probability to observe the trajectory $\{x(\cdot),y(\cdot)\}$ including its initial and final values $\{x_0,y_0\}$ and $\{x_T,y_T \}$. Here, $\bar{p}[\bar x(\cdot),\bar y(\cdot)]$ is the probability to observe the time-reversed trajectory in a time-reversed version of the process described by $\mathcal{L}(T-t)$. We set the Boltzmann constant and the temperature to unity throughout, rendering all entropies and energies dimensionless.

The entropy production defined in \eref{eqn_defJointEP} implies a fluctuation theorem of the Crooks type~\cite{Crooks1999}:
\begin{equation}\label{eqn_CFT}
\ln\frac{p(\sigma)}{\bar{p}(-\sigma)} = \sigma,
\end{equation}
where $p(\sigma)$ is the probability to obtain the entropy production $\sigma$ and $\bar{p}(-\sigma)$ denotes the probability to obtain its negative in the time-reversed process.

Additionally, \eref{eqn_CFT} implies an integral fluctuation theorem:
\begin{equation}
\left\langle e^{-\sigma} \right\rangle_{p(\sigma)} = 1.
\end{equation}
If the process starts and ends in equilibrium, we may write $\sigma = w- \Delta F$ and obtain the Jarzynski relation~\cite{Jarzynski1997}:
\begin{equation}
\left\langle e^{-w} \right\rangle_{p(w)} = e^{-\Delta F},
\end{equation}
where $w$ is the work done on the system under study and $\Delta F$ is the free energy difference between the initial and final equilibrium states.

\subsection{Coarse-graining}\label{sec:coarseGraining}
Let us now assume that we can only observe one degree of freedom $x(t)$ of the system under study. Its time evolution is obtained from \eref{eqn_jointMasterEquation} by integrating out the hidden degrees of freedom~\cite{Esposito2012}:
\begin{equation}\label{eqn_effectiveMasterEquation}
\partial_t\, p(x;t) = \tilde{\mathcal{L}}(t)\, p(x;t),
\end{equation}
with the effective generator:
\begin{equation}\label{eqn_effectiveGenerator}
\tilde{\mathcal{L}}(t) := \int dy\,\mathcal{L}(t)\,p(y|x;t),
\end{equation}
which explicitly depends on the solution of \eref{eqn_jointMasterEquation} through $p(y|x;t)$. This process of integrating out variables is known as \emph{coarse-graining}. Note that even though the marginal ensemble distribution $p(x;t) = \int dy\,p(x,y;t)$ fulfills the effective Master equation~\eref{eqn_effectiveMasterEquation}, $x(t)$ is in general not a Markov process as we will demonstrate using our model system.

\subsection{Apparent entropy production and marginal entropy production}
The effective Master equation~\eref{eqn_effectiveMasterEquation} gives rise to an effective path probability $\tilde{p}[x(\cdot)]$ with which one can define a coarse-grained~\cite{Esposito2012}, or \emph{apparent entropy production}~\cite{Mehl2012,Uhl2018} (see also the discussion in~\ref{sec:appendixA}):
\begin{equation}\label{eqn_apparentEP}
\tilde{\sigma}[x(\cdot)] := \ln\frac{\tilde{p}[x(\cdot)]}{\bar{\tilde{p}}[\bar x(\cdot)]}
\end{equation}
in analogy with \eref{eqn_defJointEP}.

Esposito~\cite{Esposito2012} showed that the coarse-graining procedure ensures that the apparent entropy production $\tilde\sigma$ on average underestimates the total entropy production $\sigma$:
\begin{equation}\label{eqn_averagesApparentEP}
\langle \sigma \rangle \geq \langle \tilde\sigma \rangle.
\end{equation}
Equality holds only when there is a separation of time scales between the dynamics of the observed degrees of freedom and the unobserved ones and if there is detailed balance between the unobserved degrees of freedom at constant observed degrees of freedom~\cite{Bo2014}. In that case the conditional distribution $p(y|x;t)$ in \eref{eqn_effectiveGenerator} can be substituted by a conditional equilibrium distribution thus rendering $y$ a bath variable. 

Concerning the fluctuations of the apparent entropy production, previous studies~\cite{Rahav2007, Mehl2012, Uhl2018, Chiang2016, Borelli2015} showed that one has to expect deviations in the fluctuation theorems. In order to see why this is the case, we contrast the apparent entropy production in \eref{eqn_apparentEP} with the \emph{marginal entropy production}:
\begin{equation}\label{eqn_defMarginalEP}
\sigma_x[x(\cdot)] := \ln\frac{p[x(\cdot)]}{\bar{p}[\bar x(\cdot)]}.
\end{equation}
Here, $p[x(\cdot)]= \int \mathcal{D}y(\cdot)\, p[x(\cdot),y(\cdot)]$ and $\bar p[\bar x(\cdot)] = \int \mathcal{D}\bar y(\cdot)\, \bar p[\bar x(\cdot),\bar y(\cdot)]$ result from appropriate marginalizations of the entire path probability of the joint process.

From its definition~\eref{eqn_defMarginalEP} it is evident that the marginal entropy production $\sigma_x$ fulfills fluctuation theorems of the usual type. Like the marginal entropy production, the apparent entropy production $\tilde{\sigma}[x(\cdot)]$ is calculated from sampled trajectories $x(t)$. However, these actually occur with relative frequencies $p[x(\cdot)]$. Therefore, we cannot expect fluctuation theorems of the usual type to hold for the apparent entropy production:
\begin{equation}
\left\langle e^{-\tilde{\sigma}[x(\cdot)]} \right\rangle_{p[x(\cdot)]} \neq \left\langle e^{-\sigma_x[x(\cdot)]} \right\rangle_{p[x(\cdot)]} = 1
\end{equation}
and
\begin{equation}
\ln\frac{p(\tilde\sigma)}{\bar{p}(-\tilde\sigma)} \neq \tilde\sigma.
\end{equation}

\section{Model system}
We consider a two-dimensional overdamped Brownian motion in a harmonic potential that is dragged through a medium at constant velocity $u$ in the $x$-direction:
\begin{eqnarray}\label{eqn_potential}
V(x,y;t) := \frac{1}{2} (x-u t)^2 + \frac{1}{2}\, y^2 - b\, (x- ut)\,y,
\end{eqnarray}
where $b$ is a coupling parameter governing the interaction between the two degrees of freedom. This model is an extension of the one-dimensional model considered by Mazonka and Jarzynski in~\cite{Mazonka1999}.

We assume that the system is initially in equilibrium with the potential $V(x,y;0)$. Experimentally, this means that the system is left alone to equilibrate before any tugging on the potential begins. After a time $T$ the driving is halted and the system is left to equilibrate. 

The two degrees of freedom shall have different mobilities $\nu_i$. We set $\nu_x$ to unity leaving us with $\nu_y := \nu$ for the $y$-dynamics. The resulting coupled overdamped Langevin equations read:
\numparts\begin{eqnarray}
\dot x &= F_x(x,y;t) + \sqrt{2}\, \xi_x(t) \label{eqn_langevinXYa} \\
\dot y &= \nu F_y(x,y;t) + \sqrt{2\nu}\, \xi_y(t), \label{eqn_langevinXYb}
\end{eqnarray}\endnumparts
with forces $F_x(x,y;t) = - \partial_x V(x,y;t)$ and $F_y(x,y;t) = - \partial_y V(x,y;t)$ and zero-mean Gaussian white noise terms $\xi_x(t)$ and $\xi_y(t)$ satisfying $\langle\xi_i(t)\,\xi_j(t')\rangle = \delta_{ij}\,\delta(t-t')$. The corresponding Fokker-Planck equation for the ensemble distribution $p(x,y;t)$ is given by: 
\begin{eqnarray}\label{eqn_FokkerPlanck}
	\partial_t p(x,y;t) = - \partial_x j_x(x, y; t) - \partial_y j_y(x, y; t) ,
\end{eqnarray}
where $j_x = (F_x - \partial_x)\,p$ and $j_y = \nu (F_y - \partial_y)\,p$ are the probability currents.

Due to the linear drift and constant diffusion coefficients in \eref{eqn_FokkerPlanck}, the solution $p(x,y;t)$ is Gaussian. According to the Langevin equations \eref{eqn_langevinXYa} and \eref{eqn_langevinXYb} the mean values obey:
\numparts\begin{eqnarray}
\dot \mu_x &= \left\langle \dot x\right\rangle = -\left(\,\mu_x - u t \right) + b\,\mu_y  \label{eqn_mux}\\
\dot \mu_y &= \left\langle \dot y\right\rangle = -\nu \,\mu_y +\nu b \left( \mu_x - ut\right). \label{eqn_muy}
\end{eqnarray}\endnumparts 
Using the Fokker-Planck equation \eref{eqn_FokkerPlanck}, the time evolution of the variance $c_{xx}$ is given by:
\begin{eqnarray}
\dot{c}_{xx} &= \int dx dy\,\partial_t p(x,y;t)\, (x^2- \mu_x^2) - 2\mu_x\,\dot{\mu}_x \nonumber \\
 &= -2\,c_{xx} + 2\,b\, c_{xy} +2 \  \label{eqn_cxx}
\end{eqnarray}
and similarly the other (co-)variances obey:
\numparts\begin{eqnarray}
\dot c_{xy} &= b \left( \nu c_{xx} + c_{yy} \right) - \left( \nu + 1 \right) c_{xy} \label{eqn_cxy} \\
\dot c_{yy} &= 2b\nu\,c_{xy}-2\nu\,c_{yy} +2\nu \label{eqn_cyy} \ .
\end{eqnarray}\endnumparts
The solution of these differential equations (with appropriate initial conditions) is then given by:
\numparts\begin{eqnarray}
	\mu_x(t) &= u t - \frac{u}{1-b^2} - \frac{u}{\nu \left( \lambda_2 - \lambda_1 \right)} \frac{\lambda_2 (1-\lambda_2) e^{-\lambda_1 t} + \lambda_1 (\lambda_1-1) e^{-\lambda_2 t}}{1-b^2} \label{eqn_meanXYsola} \\
	\mu_y(t) &= - \frac{u b}{\left( \lambda_2 - \lambda_1 \right)}  \frac{\lambda_2 (1- e^{-\lambda_1 t}) + \lambda_1 (e^{-\lambda_2 t} -1)}{1-b^2}  \label{eqn_meanXYsolb}
\end{eqnarray}\endnumparts
and 
\begin{eqnarray}\label{eqn_cij}
	c_{xx} = \frac{1}{1 - b^2}, \qquad c_{xy} = \frac{b}{1 - b^2} \qquad \mathrm{and} \qquad c_{yy} = \frac{1}{1 - b^2},
\end{eqnarray}
where the rates are specified by:
\numparts\begin{eqnarray}
	\lambda_1 &:= \frac{(1 + \nu) - \sqrt{(1 - \nu)^2 + 4 b^2 \nu}}{2} > 0 \\
	\lambda_2 &:= \frac{(1 + \nu) + \sqrt{(1 - \nu)^2 + 4 b^2 \nu}}{2} > \lambda_1>0 \ .
\end{eqnarray}\endnumparts

The degree of freedom associated with the $y$-dynamics shall be hidden from the observer who thus assumes an apparent one-dimensional motion in a dragged harmonic potential. Following \sref{sec:coarseGraining}, this effective potential $\tilde{V}(x;t)$ is obtained by marginalizing \eref{eqn_FokkerPlanck}:
\begin{eqnarray}\label{eqn_marginalFokkerPlanck}
	\partial_t p(x;t) = - \partial_x \bigg ( \underbrace{\int dy F_x(x, y; t) p(y|x;t)}_{=: - \partial_x \tilde{V}(x;t)}  - \partial_x \bigg ) p(x;t)
\end{eqnarray}
yielding:
\begin{eqnarray}\label{eqn_coarseGrainedPotential}
	\tilde{V}(x;t) &= \frac{1-b^2}{2} (x-ut)^2 + u b^2 (x-ut) \frac{e^{-\lambda_2 t} - e^{-\lambda_1 t}}{\lambda_2 - \lambda_1} \nonumber \\
					&\hspace{-2.5mm} \overset{\lambda_1 t \, \gg \, 1}{\longrightarrow}  \frac{1-b^2}{2} (x-ut)^2  \ .
\end{eqnarray}
Therefore, an experimenter unaware of the second degree of freedom would use the potential in \eref{eqn_coarseGrainedPotential} to model the system. This is because, experimentally, one would fit the potential to the observed initial equilibrium distribution, which is given by:
\begin{eqnarray}
\int dy\, \exp \left( -V(x,y;0) \right) \propto  \exp\left( - \frac{1-b^2}{2} x^2 \right).
\end{eqnarray}

\subsection{Work distribution}
The fluctuating total work $w$ done on the system is identified following the standard prescription of stochastic energetics \cite{Sekimoto1998}:
\begin{eqnarray}\label{eqn_totalwork}
	w[x(\cdot),y(\cdot)] &= \int \limits_0^T dt\,\partial_t V(x,y;t) \nonumber \\
	  &= \int \limits_0^T dt \left[- u (x-ut) + u b y \right] \ .
\end{eqnarray}
It equals the entropy production $\sigma$ since the free energy of the system remains constant during the process. Accordingly, the apparent work $\tilde{w}$ is given by
\begin{eqnarray}\label{eqn_langevinApparentWork}
\tilde{w}[x(\cdot)] &= \int \limits_0^T dt\,\partial_t \tilde{V}(x;t) \nonumber \\
					&= -\int \limits_0^T dt\,u (1-b^2) \,(x-u t) .
\end{eqnarray}
Together with \eref{eqn_langevinXYa} and \eref{eqn_langevinXYb} this specifies a system of three linearly coupled Langevin equations. From now on we switch to the moving reference frame $x \rightarrow x-u t$. The coupled system of Langevin equations then reads:
\numparts\begin{eqnarray}
\dot x &= -x + b\,y - u + \sqrt{2}\, \xi_x(t) \label{eqn_langevinXYWa}  \\
\dot y &= -\nu\,y +\nu b\,x + \sqrt{2\nu}\, \xi_y(t) \label{eqn_langevinXYWb} \\
\dot{\tilde{w}} &= -u (1-b^2) x, \ \label{eqn_langevinXYWc}
\end{eqnarray}\endnumparts
with the corresponding Fokker-Planck equation for this joint process:
\begin{eqnarray}\label{eqn_FokkerPlanckXYW}
\partial_t p(x,y,\tilde{w};t) = &\left[\partial_x\left( x - by + u\right) + \partial^2_x + \partial_y\left( \nu y - \nu b x\right) \right. \nonumber \\
			& \qquad \qquad  \left. + \nu \partial_y^2 +u (1-b^2) x\,\partial_{\tilde w} \right]\,p(x,y,\tilde{w};t) \ .
\end{eqnarray}
The solution is again Gaussian and the moments are obtained in the same way as before yielding the following asymptotic expression for the mean apparent work 
\begin{eqnarray}\label{eqn_asymptMean}
\mu_{\tilde{w}}(T) &\overset{\lambda_1 T \, \gg \, 1}{\longrightarrow} u^2 T - u^2 \frac{b^2+\nu}{\nu(1-b^2)}
\end{eqnarray}
and for the variance 
\begin{eqnarray}\label{eqn_asymptVar}
 c_{\tilde{w} \tilde{w}}(T) &\overset{\lambda_1 T \, \gg \, 1}{\longrightarrow} 2 u^2 \frac{b^2+\nu}{\nu} T - 2 u^2 \frac{b^2 + 2 b^2 \nu + \nu^2}{\nu^2 (1-b^2)}  .
\end{eqnarray}

\begin{figure}[h!]
	\begin{flushright}
    \includegraphics[width = 0.425 \textwidth]{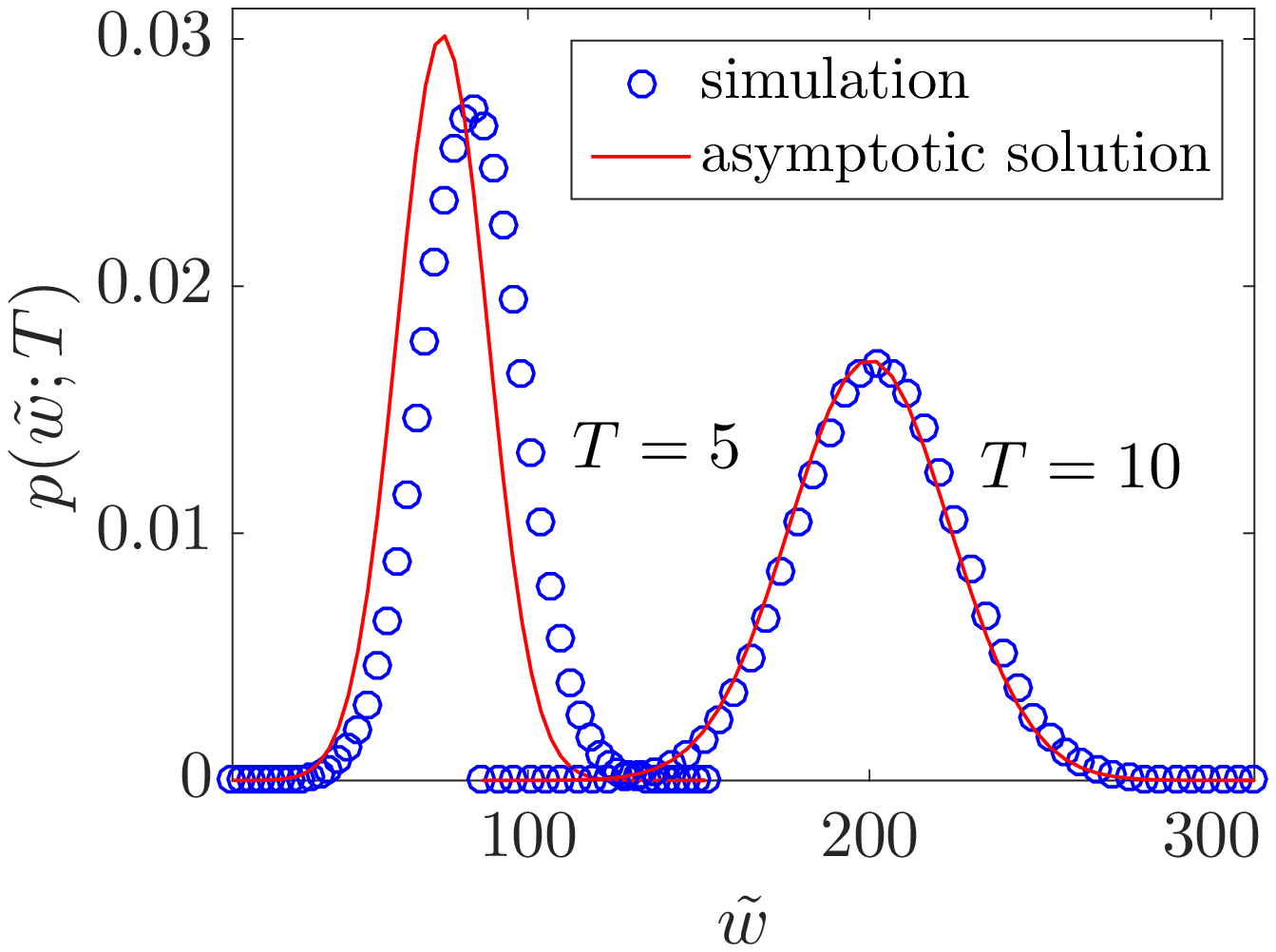}%
    \includegraphics[width = 0.425 \textwidth]{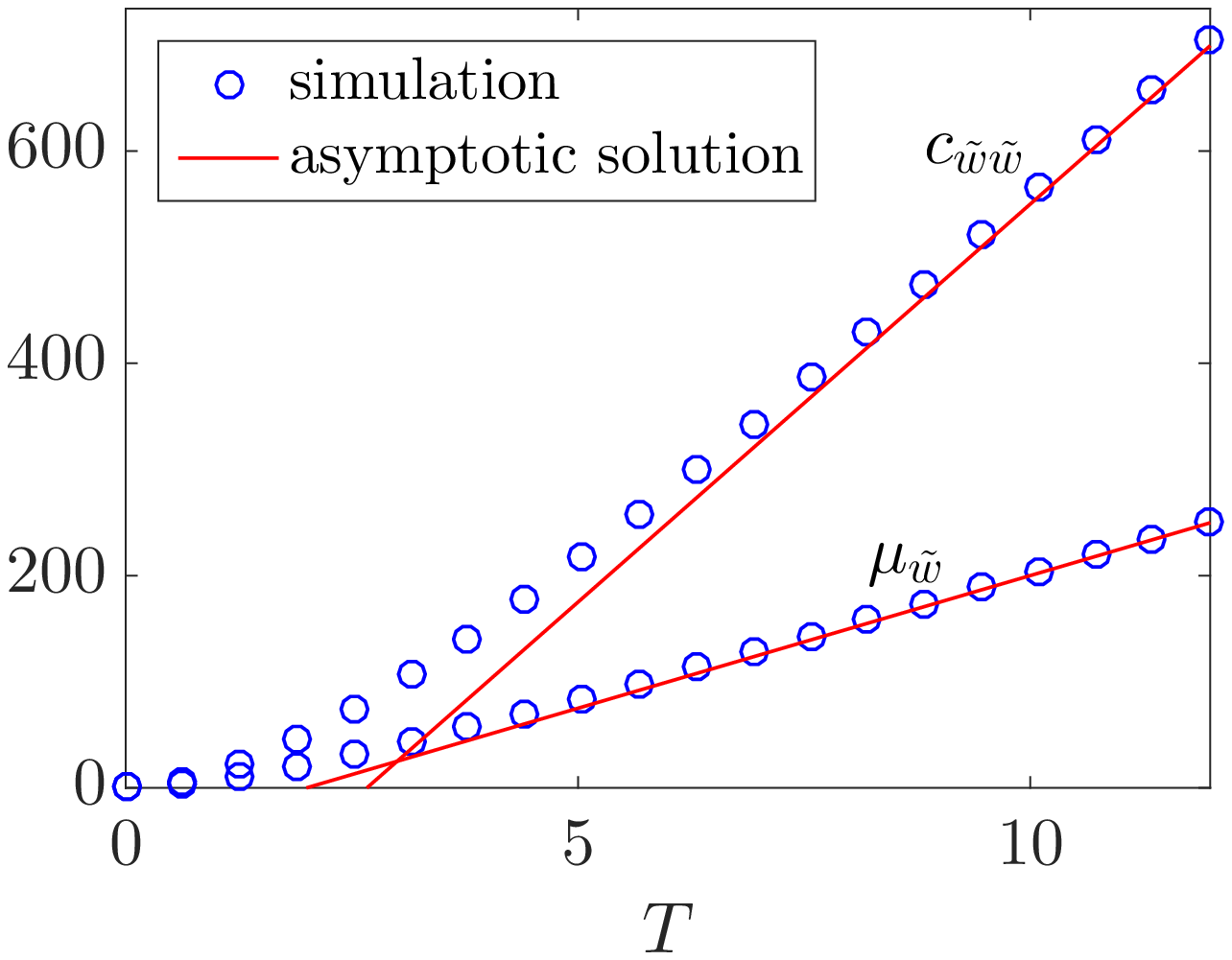}
    \end{flushright}\vspace{-6mm}
  \caption{Left: Histograms for the apparent work $\tilde{w}$ gained from simulations (time step: $dt = 10^{-2}$ and $10^6$ realizations) of the discretized Langevin equation~\eref{eqn_langevinXYWa}-\eref{eqn_langevinXYWc} for model parameters $u=5$, $b=1/2$ and $\nu=1/2$. Additionally, the asymptotic solutions are shown. Right: Comparison of simulations and asymptotic solution for the mean and the variance of the apparent work.}
\label{fig:work distribution}
\end{figure}

The asymptotic distribution $p(\tilde{w};T)$ is shown in \fref{fig:work distribution} for a representative set of parameters. We also show results from numerically evaluating \eref{eqn_langevinXYWc} using $x$-trajectories obtained from simulating the joint system in~\eref{eqn_langevinXYWa} and~\eref{eqn_langevinXYWb}. We note that the histograms obtained from simulations asymptotically converge to the distribution specified by~\eref{eqn_asymptMean} and~\eref{eqn_asymptVar}.

\subsection{Fluctuation Theorems}
Our process starts and ends in equilibrium with $\Delta F = 0$. The work given in~\eref{eqn_totalwork} is invariant under time reversal $t\rightarrow T-t$. Thus:
\begin{eqnarray}
p(\sigma) = p(w) = \bar{p}(w).
\end{eqnarray}
The work distribution therefore fulfills a detailed fluctuation theorem~\cite{Seifert2012}:
\begin{eqnarray}\label{eqn_DFT}
\ln\frac{p(w)}{p(-w)} = w.
\end{eqnarray}
In contrast, the detailed fluctuation theorem for the apparent work $\tilde{w}$ reads:
\begin{eqnarray}
	\ln \frac{p(\tilde{w})}{p(-\tilde{w})} &= \frac{2\mu_{\tilde{w}}}{c_{\tilde{w}\tilde{w}}} \tilde{w} \nonumber \\
	&\hspace{-2.5mm} \overset{\lambda_1 T \, \gg \, 1}{\longrightarrow} \frac{\nu}{\nu+b^2} \tilde{w}.\label{eqn_dftWtildeAsympt}
\end{eqnarray}
Since $p(\tilde{w})$ is Gaussian in our model, deviations from the usual detailed fluctuation theorem only manifest themselves in an altered slope. 

\begin{figure}[H]
	\begin{flushright}
    \includegraphics[width = 0.425 \textwidth]{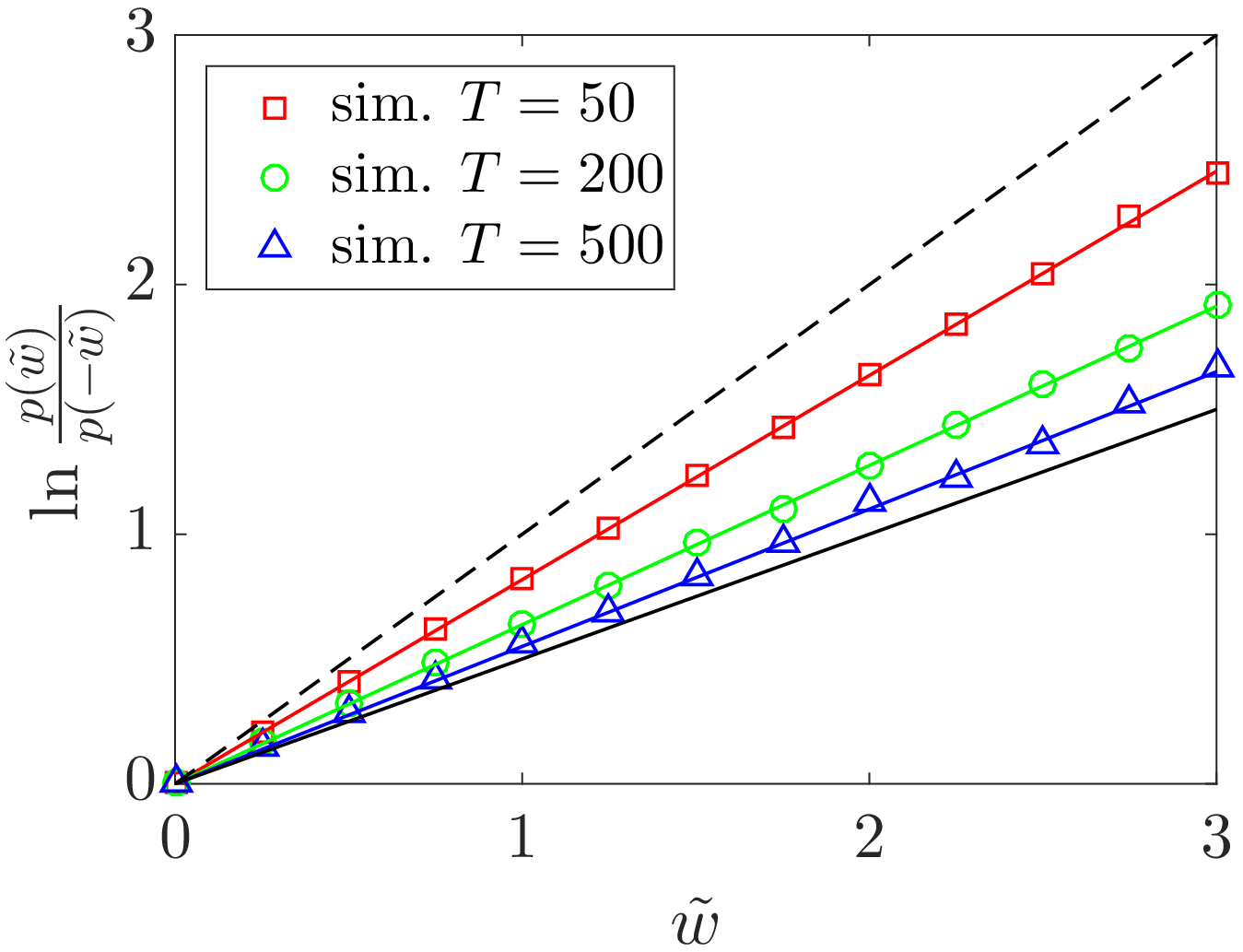}
    \includegraphics[width = 0.425 \textwidth]{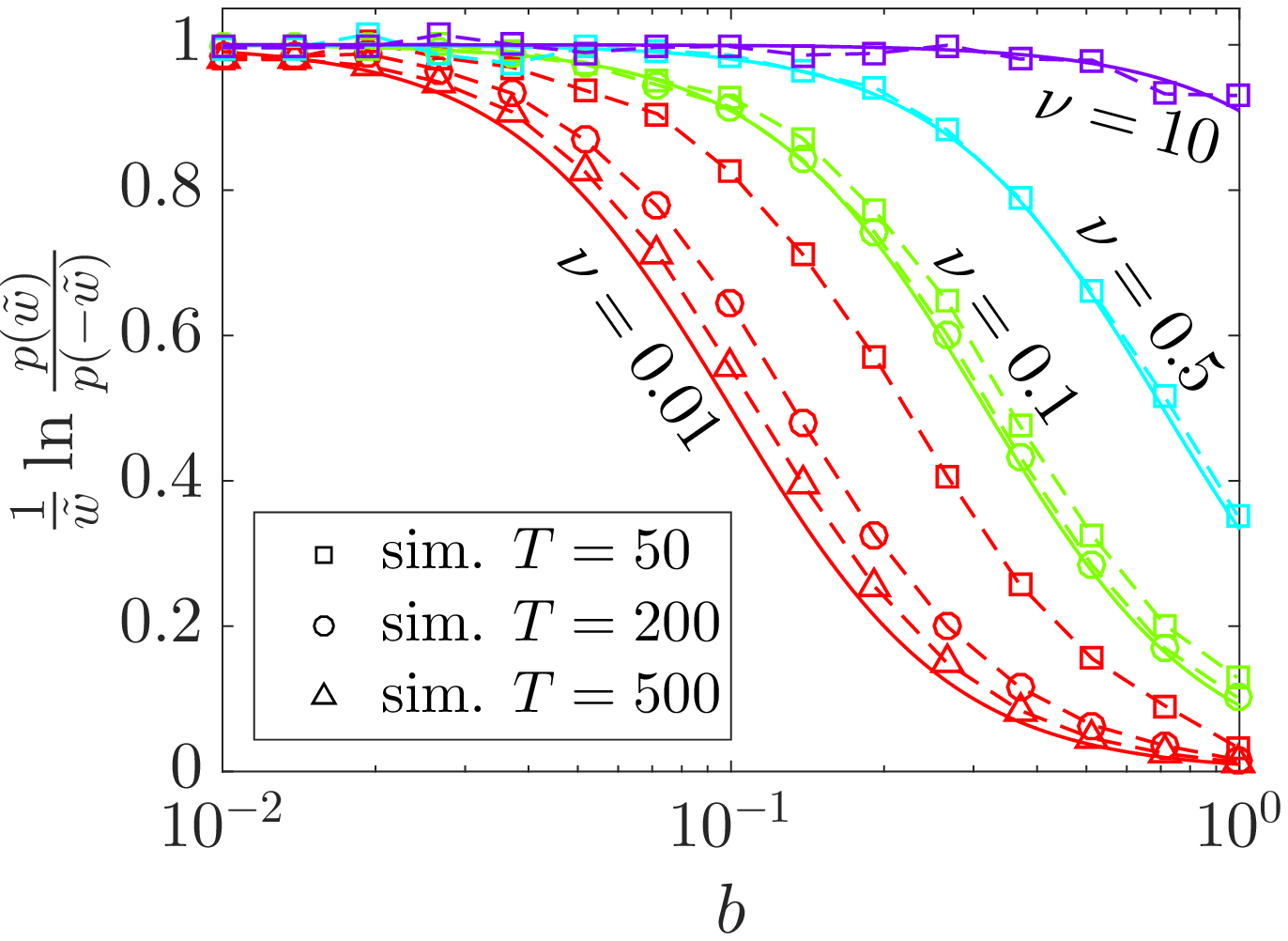}
    \end{flushright}\vspace{-6mm} 
  \caption{Left: Asymptotic detailed fluctuation theorem (black solid line) together with the curves obtained from the histograms of apparent work values gained from simulations (time step: $dt = 10^{-1}$ and $10^6$ realizations) of different lengths (symbols). The parameters are $u = 0.1$, $b=0.1$ and $\nu=0.01$. The simulation results are accompanied by linear fits. Error bars are smaller than the symbol size. The dashed line indicates a slope of one. We see that for large times the simulation results converge to the asymptotic fluctuation theorem. Right: Slope of the detailed fluctuation theorem~\eref{eqn_dftWtildeAsympt} in dependence of $b$ and $\nu$ together with simulations for $u = 0.1$ (time step: $dt = 10^{-1}$ and $10^5$ realizations). We infer that in the absence of coupling ($b\rightarrow 0$) and when there is a separation of time-scales (large $\nu$) one recovers the original fluctuation relation with slope one.}\label{fig:cft}
\end{figure}

\Fref{fig:cft} shows the asymptotic detailed fluctuation theorem given by \eref{eqn_dftWtildeAsympt} together with the fluctuation theorems calculated from the histograms of the apparent work obtained from simulations. 

In the limiting cases of no coupling ($b \rightarrow 0$) and time scale separation ($\nu \rightarrow \infty$) the detailed fluctuation theorem is fulfilled in the usual form. In the former case the hidden variable $y$ decouples from the observed variable $x$ and thus the apparent work \eref{eqn_langevinApparentWork} equals the total work \eref{eqn_totalwork}. In the latter case the hidden degree of freedom is pushed into a conditional equilibrium with the observed variable. In this situation coarse-graining delivers a thermodynamically consistent description of the observed process.

For completeness, we also state the asymptotic integral fluctuation theorem:
\begin{eqnarray}\label{eqn_iftWtildeAsympt}
	\langle e^{-\tilde{w}} \rangle &= \exp\left( \frac{c_{\tilde{w}\tilde{w}}}{2} - \mu_{\tilde{w}} \right) \nonumber  \\
								   &\hspace{-2.5mm} \overset{\lambda_1 T \, \gg \, 1}{\longrightarrow} \exp\left( \frac{u^2 b^2 T}{\nu} - \frac{u^2 b^2 (\nu+1)}{\nu^2 \left( 1 - b^2 \right)} \right) 	.
\end{eqnarray}
Again, for $b\rightarrow 0$ and $\nu\rightarrow\infty$ the fluctuation theorem holds in the usual form.

\subsection{Marginal fluctuation theorem}
Having established that using the apparent entropy production causes deviations in fluctuation relations, we now calculate the marginal entropy production $\sigma_x$ defined in~\eref{eqn_defMarginalEP}. 

Let us consider the problem of calculating the marginal path probability in general. Instead of calculating it directly, it is instructive to see how the marginal entropy production emerges from the fluctuation relation for the total entropy production:
\begin{equation}
\left\langle e^{-\sigma[x(\cdot),y(\cdot)]} \right\rangle_{p[x(\cdot),y(\cdot)]} = \left\langle \left\langle e^{-\sigma[x(\cdot),y(\cdot)]} \right\rangle_{p[y(\cdot)|x(\cdot)]} \right\rangle_{p[x(\cdot)]}.
\end{equation}
The inner average reads:
\begin{eqnarray}
\left\langle e^{-\sigma[x(\cdot),y(\cdot)]} \right\rangle_{p[y(\cdot)|x(\cdot)]} &= \int \mathcal{D}y(\cdot)\, p[y(\cdot)|x(\cdot)]\, \frac{ \bar p[\bar x(\cdot),\bar y(\cdot)]}{p[x(\cdot),y(\cdot)]}\nonumber\\
&= \frac{\int \mathcal{D}y(\cdot)\,\bar p[\bar x(\cdot),\bar y(\cdot)]}{p[x(\cdot)]} \nonumber\\
&= \frac{\bar p[\bar x(\cdot)]}{p[x(\cdot)]}\nonumber\\
&\overset{(10)}{=} e^{-\sigma_x},\label{eqn_marginalEntProdFromCondAvg}
\end{eqnarray}
such that:
\begin{equation}
\left\langle e^{-\sigma_x[x(\cdot)]} \right\rangle_{p[x(\cdot)]} = 1.
\end{equation}

In our model we identify the marginal entropy production in \eref{eqn_marginalEntProdFromCondAvg} as the \emph{marginal work} $w_x$:
\begin{eqnarray}
e^{-w_x} &:= \left\langle e^{-w[x(\cdot),y(\cdot)]} \right\rangle_{p[y(\cdot)|x(\cdot)]} \nonumber\\
&= \frac{\int \mathcal{D}y(\cdot)\, e^{-w}\, p[x(\cdot),y(\cdot)]}{\int \mathcal{D}y(\cdot)\, p[x(\cdot),y(\cdot)]}\nonumber\\
&=: \frac{I_1}{I_0},
\end{eqnarray}
with
\begin{equation}\label{eqn_defIAlpha}
\fl I_{\alpha} := \int dy_0\int dy_T\, p(x_0,y_0) \int\mathcal{D}y(\cdot)\,p[x(\cdot),y(\cdot)|x_0,y_0] \,\exp{\left\{ \alpha \int\limits_0^T dt \left(u x-u b y\right) \right\}},
\end{equation}
where we now explicitly indicated the integration over the boundary terms.

The trajectory probability follows from~\eref{eqn_langevinXYWa} and~\eref{eqn_langevinXYWb} and is up to normalization given by:
\begin{equation}
\fl p[x(\cdot),y(\cdot)|x_0,y_0] \propto \exp\left\{ -\frac{1}{4} \int\limits_0^T dt \left( \dot{x} + x - b y + u \right)^2  -\frac{1}{4\nu} \int\limits_0^T dt \left( \dot{y} + \nu y - \nu b x \right)^2   \right\}.
\end{equation}
Since the process starts in equilibrium, the initial condition reads:
\begin{equation}
p(x_0,y_0) \propto \exp\left\{ -\frac{1}{2}\,x_0^2-\frac{1}{2}\,y_0^2 + b\,x_0\,y_0 \right\}.
\end{equation}
Thus we can write:
\begin{equation}
I_{\alpha} \propto \int dy_0\int  dy_T\int\limits_{(y_0,0)}^{(y_T,T)}\mathcal{D}y(\cdot)\exp \left\{-\frac{1}{4} S_\alpha\left[ x(\cdot),y(\cdot),x_0,y_0\right ]\right\},
\end{equation}
where:
\begin{eqnarray} \label{eqn_action}
\fl S_\alpha := \int\limits_0^T dt \left( \dot{x} + x - b y + u \right)^2 +\frac{1}{\nu} \int\limits_0^T dt \left( \dot{y} + \nu\,y - \nu b\, x \right)^2 -4 \alpha \int\limits_0^T dt \left(u x-u b y\right)\nonumber\\
+2 x_0^2 +2 y_0^2-4 b x_0 y_0.
\end{eqnarray}
In $S_\alpha$ all terms that do not depend on $\alpha$, $y(\cdot)$, $y_0$ or $y_T$ need not be considered since they will cancel upon taking the ratio $I_1/I_0$. Thus, after partial integration:
\begin{eqnarray}
\fl S_\alpha = \int\limits_0^T dt \left[ y^2 (b^2+\nu)-2 b\, y\, (x+u+\nu\, x -2 u \alpha) +\frac{1}{\nu}\,\dot{y}^2 -4 \alpha u x \right]\nonumber\\
+ y_0^2+y_T^2 -2b (x_0 y_0- x_T y_T) + \mathrm{const.}
\end{eqnarray}
The path integral $I_\alpha$ is Gaussian and can therefore be calculated with the saddle-point method, i.e. we expand $S_\alpha$ around its extremum $\bar{S}_\alpha := S_\alpha\left[x(\cdot),\bar y(\cdot)\right]$. We set $y(\cdot) := \bar y(\cdot) + \delta(\cdot)$ and obtain after partial integration:
\begin{eqnarray}
\fl S_\alpha = \bar S_\alpha + \delta_0^2 + \delta_T^2 + \int\limits_0^T dt \left[(b^2+\nu) \delta^2+ \frac{1}{\nu}\dot{\delta}^2 \right] \nonumber\\
+  \delta_0 \left( -2b x_0 + 2 \bar y_0 - \frac{2}{\nu}\,\dot{\bar{y}}|_0 \right)+ \delta_T \left( -2b x_T + 2 \bar y_T + \frac{2}{\nu}\,\dot{\bar{y}}|_T \right)\nonumber\\
+ \int\limits_0^T dt \,\delta\left[-\frac{2}{\nu} \ddot{\bar{y}} + 2(b^2+\nu) \bar{y}- 2b (x+u+\nu x - 2 \alpha u)\right] + \mathrm{const.}
\end{eqnarray}
The extremal trajectory $\bar{y}(\cdot)$ is thus obtained by solving the Euler-Lagrange equation:  
\begin{equation}\label{eqn_ELE}
\ddot{\bar{y}} -a^2 \bar{y} = -c\, x(t) + (2 \alpha-1)\,  \nu b u, 
\end{equation}
where:
\begin{eqnarray}
a := \sqrt{ \nu(\nu+b^2)}\label{eqn_defa}\\
c := \nu b\,(1+\nu) \label{eqn_defc}.
\end{eqnarray}
The solution needs to obey the boundary conditions:
\begin{eqnarray}
0 = \dot{\bar{y}}|_0 - \nu\,\bar{y}_0 + \nu b \, x_0\label{eqn_boundary0}  \\
0 = \dot{\bar{y}}|_T + \nu\,\bar{y}_T - \nu b \, x_T.\label{eqn_boundaryT} 
\end{eqnarray}
It is given by:
\begin{eqnarray}
\fl \bar{y}_\alpha(t) = \frac{c}{a}\, A(t) - \frac{2 \alpha-1}{a^2}\, \nu b u\left[ 1- \frac{g(t) + g(T-t)}{a} \right] + \frac{c\,g(t)}{a \nu}\left[ B(T)- \frac{\nu}{a} A(T) \right]\nonumber\\
+ \frac{b}{a}\left[ x_0\, g(T-t) + x_T\, g(t)\right],
\end{eqnarray}
where 
\begin{eqnarray}
A(t) := \int\limits_0^t dt \, x(t')\,\sinh{a(t'-t)},\\
B(t) := \int\limits_0^t dt \, x(t')\,\cosh{a(t'-t)}
\end{eqnarray}
and
\begin{equation}
g(t) := \frac{\frac{a^2}{\nu} \cosh{at}+a \sinh{at}}{\left( 1+\frac{a^2}{\nu^2}\right) \sinh{aT} + \frac{2 a}{\nu}\cosh{aT}  }.
\end{equation}
The remaining path integral over $\delta(\cdot)$ need not be carried out since it does not depend on $\alpha$ and cancels when taking the ratio $I_1/I_0$. We therefore find after partial integration and using~\eref{eqn_ELE},~\eref{eqn_defa},~\eref{eqn_defc},~\eref{eqn_boundary0}~and~\eref{eqn_boundaryT}:
\begin{eqnarray}
\hspace{-15mm}S_\alpha &= \bar{S}_\alpha + \mathrm{const.} \nonumber\\
\hspace{-15mm}&= -\frac{1}{\nu} \int\limits_0^T dt \, \bar{y}\left[c x + (1-2\alpha) \nu b u \right] - 4 \alpha u \int\limits_0^T dt\, x - b\left(x_T \bar{y}_T+ x_0 \, \bar{y}_0\right) + \mathrm{const.}
\end{eqnarray}
With this we obtain:
\begin{eqnarray}
\hspace{-20mm} \bar S_1 - \bar S_0 = -\frac{1}{\nu} \int\limits_0^T dt \, \Delta(t)\left(c x + \nu b u \right) + 2 b u \int\limits_0^T dt\,\bar{y}_1 - 4 u \int\limits_0^T dt\,x - b \Delta(0) \,(x_0+x_T),
\end{eqnarray}
where
\begin{eqnarray}
\Delta(t) &:= y_1(t)-y_0(t)\nonumber\\
&= -\frac{2 \nu b u}{a^2}\left[ 1- \frac{g(t) + g(T-t)}{a} \right].
\end{eqnarray}
We finally obtain the marginal work $w_x$:
\begin{eqnarray}
\fl w_x[x(\cdot)]  &= \ln{I_0} - \ln{I_1}\nonumber\\
\hspace{-15mm} &= \frac{\bar S_1-\bar S_0}{4}\nonumber\\
\hspace{-15mm} &=-u\,(1-b^2)\,\frac{\nu}{\nu+b^2}\int\limits_0^T dt\, x + \frac{b^2 u}{a\,} G(a,\nu, T)\, (x_0+x_T) \nonumber\\
\hspace{-15mm}&\qquad  - b^2  u \frac{\nu+1}{\nu+b^2}\,\left(1-\frac{a}{\nu}\,G(a,\nu, T)\right) \int\limits_0^T dt\,x(t)\,\left( \frac{\sinh{at}+\sinh{a(T-t)}}{\sinh{aT}} \right), \label{eqn_wx}
\end{eqnarray}
with:
\begin{eqnarray}
G(a,\nu,T) := \int\limits_0^T dt\,g(t) = \frac{1}{\frac{a}{\nu}+\coth{\frac{a T}{2}}}.
\end{eqnarray}

\begin{figure}[H]
	\begin{flushright}
    \includegraphics[width = 0.9 \textwidth]{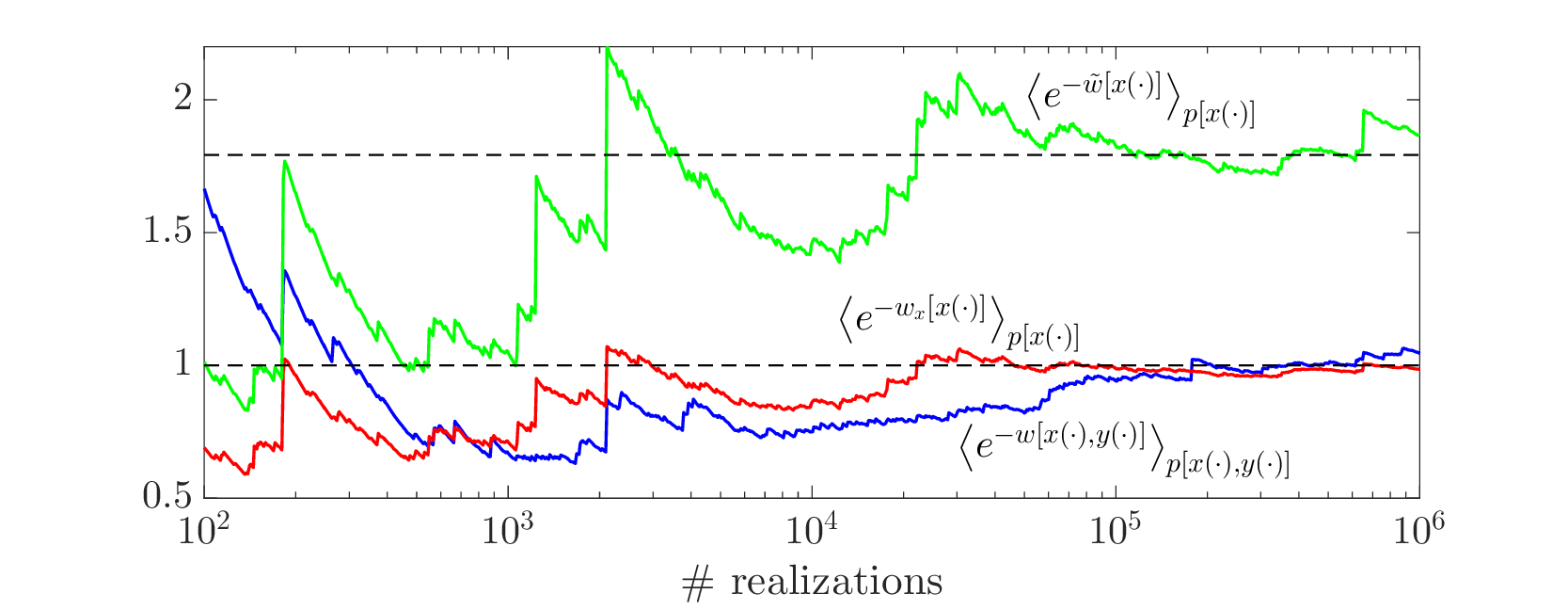}
    \end{flushright}\vspace{-6mm} 
  \caption{Convergence of the integral fluctuation theorem for the total work $w$ calculated from numerically simulated trajectories of both degrees of freedom (blue line), the marginal work $w_x$ (red line) and the apparent work $\tilde{w}$ (green line), which have been calculated using the $x$-trajectory alone. The upper dashed line indicates the time asymptotic fluctuation theorem \eref{eqn_iftWtildeAsympt} for the apparent work. The simulation parameters are: $u = 1$, $b = 0.5$, $\nu = 1$, $T=5$ and time step $dt = 10^{-2}$.}\label{fig:marginal_ift}
\end{figure}

\Fref{fig:marginal_ift} shows the convergence of the integral fluctuation theorem for the marginal work $w_x$ calculated from the $x$-trajectories of the simulation. This is contrasted with the fluctuation theorem for the apparent work $\tilde{w}$ calculated from \eref{eqn_langevinApparentWork}. Additionally, we used both degrees of freedom to calculate the total work $w$ using \eref{eqn_totalwork} for which we also plotted the integral fluctuation theorem.

As expected the apparent work $\tilde{w}$ does not fulfill the integral fluctuation theorem while both the total and marginal work do. Interestingly, the convergence is faster for the marginal work $w_x$ than for total work $w$, since part of the averaging has already been accomplished by integrating out the $y$-variable.

We now turn to some limiting cases. When there is no coupling, i.e. for $b\rightarrow 0$, the marginal work converges to the apparent work and both converge to the total work given in \eref{eqn_totalwork}:
\begin{equation}
\lim_{b\rightarrow 0} w_x  = \lim_{b\rightarrow0}\tilde{w} = \lim_{b\rightarrow0} w =- u \int\limits_0^T dt\,x(t) .
\end{equation}

The limit $\nu \rightarrow \infty$, i.e. when there is a separation of time scales, is more intricate: We find from \eref{eqn_defa} that $a \rightarrow \nu$. Thus also $a \rightarrow \infty$, which implies $G(a,\nu, T) \rightarrow 1/2$. Additionally, the last integral in \eref{eqn_wx} vanishes leaving us with:
\begin{equation}
\lim_{\nu\rightarrow \infty} w_x = \tilde{w}= -u\,(1-b^2)\,\int\limits_0^T dt\, x(t) .
\end{equation}
This result is in agreement with our expectation that coarse-graining delivers a consistent description when there is a separation of time scales.

Lastly, for large $T$ we find: $G(a,\nu, T) \rightarrow \nu/(a+\nu)$, which implies that the first term of \eref{eqn_wx} grows linearly with $T$ while the others stay roughly constant. We may thus neglect the second and third terms leaving us with:
\begin{equation}
w_x[x(\cdot)] \overset{a T \gg 1}{\approx} -u\,(1-b^2)\,\frac{\nu}{\nu+b^2}\int\limits_0^T dt\, x(t).
\end{equation}
This is an interesting result, since with \eref{eqn_langevinApparentWork} it means that asymptotically:
\begin{equation}\label{eqn_relationWxW}
w_x  = \frac{\nu}{\nu+b^2}\, \tilde{w},
\end{equation}
which immediately implies the asymptotic detailed fluctuation theorem for $\tilde{w}$:
\begin{equation}
\ln\frac{p_{w_x}(w_x)}{p_{w_x}(-w_x)} = w_x \overset{(63)}{\Longleftrightarrow} \ln\frac{p_{\tilde{w}}(\tilde{w})}{p_{\tilde{w}}(-\tilde{w})} =   \frac{\nu}{\nu+b^2}\, \tilde{w},
\end{equation}
in agreement with our previous result in \eref{eqn_dftWtildeAsympt}.

\section{Discussion}
We demonstrated that the fluctuation theorems for the apparent entropy production of coarse-grained systems deviate from their usual form.

Naively, this is unexpected since there is an effective description of the marginal process $x(t)$ with the effective Master equation~\eref{eqn_marginalFokkerPlanck}. Yet, this effective description is only valid on the ensemble level and not correct on the trajectory level. The apparent entropy production expects the marginal process to be Markovian because it is defined using the effective Master equation. However, $x(t)$ is not a Markov process as we demonstrate in~\ref{sec:appendixB} by means of the Chapman-Kolmogorov equation. The Master equation for $p(x;t)$ governs only the ensemble level. It does not imply that $x(t)$ is a Markov process. For this it would also have to hold true for any transition probabilities $p(x;t|x_0;t_0)$~\cite{VanKampen2007}.

This also explains why there are no deviations in the fluctuation theorems for the apparent entropy production in the limit of decoupling ($b\rightarrow0$) and time-scale separation ($\nu \rightarrow \infty$), since in these cases $x(t)$ becomes Markovian.

There is the possibility of defining a marginal entropy production which is based on the trajectory level. It therefore covers the entire statistics of the marginal process. Naturally, fluctuation theorems for this quantity hold in their usual form.

We point out that our findings depend on the long-time limit we performed. This is because we neglect the relaxation terms in the coarse-grained potential in~\eref{eqn_coarseGrainedPotential}, which are due to the coarse-graining scheme capturing the relaxation of the hidden degree of freedom. These terms produce an additional contribution to the apparent work in~\eref{eqn_langevinApparentWork} which does not grow with $T$ and therefore does not affect the asymptotic detailed fluctuation theorem. Similarly, coarse-graining produces relaxation terms after the driving has stopped, which can be neglected as well.

The finding that the apparent entropy production does not generally fulfill the standard fluctuation theorems is in agreement with~\cite{Rahav2007,Mehl2012,Uhl2018,Chiang2016,Borelli2015}. Because in our model the work distribution is Gaussian, the detailed fluctuation theorem remains linear with a modified slope. However, for other setups there can be a distinctly nonlinear behavior.~\cite{Uhl2018}.

Our results suggest that one can use fluctuation theorems to infer the existence of hidden degrees of freedom: Imagine an experimenter only having access to one degree of freedom. They would model the process with an effective description valid on the ensemble level. Subsequently, the fluctuation theorem for the apparent entropy can be employed and will reveal the existence of hidden degrees of freedom. Furthermore, with a suitable model of all degrees of freedom at hand, one could infer model parameters from the deviations in the fluctuation theorem.

Arguably, the most complete entropy production for one of several degrees of freedom is the marginal entropy production and one should strive to use it, although it might be hard to calculate in practical applications.

\section{Conclusion}
In this article we considered an analytically solvable model of a stochastic system with a visible and a hidden slow degree of freedom. For this model we studied the fluctuations of the apparent entropy production which is defined on the basis of a coarse-grained effective description. We were able to predict deviations in the fluctuation theorem. The reason for these deviations lies in the non-Markovianity of the visible process, which is not captured by the coarse-grained description. We proposed as an alternative the marginal entropy production for which the fluctuation theorem naturally holds.

\section*{Acknowledgments}
We thank Andreas Engel for fruitful discussions and a critical reading of the manuscript. M.K. acknowledges financial support from the Heinz-Neum\"uller Foundation.

\section*{Author contribution statement}
Both authors contributed equally to this paper.

\appendix
\section{Coarse-graining scheme on the basis of mean local velocities}\label{sec:appendixA}
Here, we want to show that the coarse-graining scheme used in~\cite{Mehl2012} is consistent with our definition of coarse-graining.

In~\cite{Mehl2012} the authors follow the usual definition of stochastic entropy production~\cite{Seifert2005} and identify the apparent entropy production as the product of effective mean velocity $\tilde{v}(x;t)$ of the observed degree of freedom times the velocity $\dot{x}$:
\begin{equation}
\tilde{\sigma} := \int\limits_0^T dt\, \tilde{v}(x;t) \cdot \dot{x}(t),
\end{equation}
where the effective mean velocity is defined by:
\begin{equation}\label{eqn_effectiveMeanVelocity}
\tilde{v}(x;t) := \int dy\, v_x(x,y;t)\,p(y|x;t).
\end{equation}
The mean velocities are given by:
\begin{eqnarray}
v_x(x,y;t) = \nu_x\, F_x(x,y;t) - D_x\,\frac{\partial_x p(x,y;t)}{p(x,y;t)}\label{eqn_defMeanVelocityX}\\
v_y(x,y;t) = \nu_y\, F_y(x,y;t) - D_y\,\frac{\partial_y p(x,y;t)}{p(x,y;t)},\label{eqn_defMeanVelocityY}
\end{eqnarray}
where the $\nu_i$ are the mobilities and the $D_i$ the diffusivities of the individual degrees of freedom and the $F_i$ are the forces acting upon them.

We recover this definition using our coarse-graining scheme outlined in \sref{sec:coarseGraining} by realizing that $\mathcal{L}(t)$ is the Fokker-Planck operator:
\begin{equation}
\mathcal{L}(t) = -\partial_x (\nu_x\,F_x(x,y;t)-D_x\,\partial_x) -\partial_y (\nu_y\,F_y(x,y;t)-D_y\,\partial_y).
\end{equation}
According to \eref{eqn_effectiveGenerator} and after partial integration, the effective operator is then given by:
\begin{equation}
\tilde\mathcal{L}(t) = -\partial_x (\nu_x\,\tilde{F}_x(x;t)-D_x\,\partial_x),
\end{equation}
with the effective force:
\begin{equation}
\tilde{F}_x(x;t) := \int dy \, F_x(x,y;t)\, p(y|x;t).
\end{equation}
With this we define an effective mean local velocity $\tilde{v}(x;t)$ in accordance with~\eref{eqn_defMeanVelocityX} and~\eref{eqn_defMeanVelocityY}:
\begin{eqnarray}
\tilde{v}(x;t) &:= \nu_x\, \tilde{F}_x(x;t) - D_x\,\frac{\partial_x p(x;t)}{p(x;t)}\nonumber\\
 &= \int dy\,\left[\nu_x\, F_x(x,y;t)\, p(y|x;t) - D_x\,\frac{\partial_x p(x,y;t)}{p(x;t)}\right]\nonumber\\
 &= \int dy\,\left[\nu_x\, F_x(x,y;t) - D_x\,\frac{\partial_x p(x,y;t)}{p(x,y;t)}\right]\, p(y|x;t)\nonumber\\
 &= \int dy\,v_x(x,y;t)\, p(y|x;t),
\end{eqnarray}
which agrees with \eref{eqn_effectiveMeanVelocity}.

\section{The coarse-grained process is in general not Markovian}\label{sec:appendixB}
Even though the marginal distribution $p(x;t)$ of the coarse-grained process fulfills the effective Fokker-Planck equation~\eref{eqn_marginalFokkerPlanck}, it is not a Markov process. We prove this by showing that the Chapman-Kolmogorov equation is not fulfilled:
\begin{equation}\label{eqn_chapman_kolmogorov}
p(x_2;t_2|x_0;0) \neq \int dx_1\,p(x_2;t_2|x_1;t_1)\,p(x_1;t_1|x_0;0),
\end{equation}
where $t_2> t_1 > 0$.

For this we require the propagator of the marginal process:
\begin{eqnarray}
p(x;t|x_0;0) = \int dy \int dy_0\, p(x,y;t|x_0,y_0;0)\, p(y_0|x_0;0).
\end{eqnarray}
The propagator of the joint process $p(x,y;t|x_0,y_0;0)$ can be calculated from the solution of the Fokker-Planck equation~\eref{eqn_FokkerPlanck} with a delta-like initial distribution and $p(y_0|x_0;0)$ follows from the initial equilibrium distribution. The propagator $p(x;t|x_0;0)$ is Gaussian. The expressions for its mean and the variance are too long to be displayed here.

\begin{figure}[H]
	\begin{flushright}
    \includegraphics[width = 0.9 \textwidth]{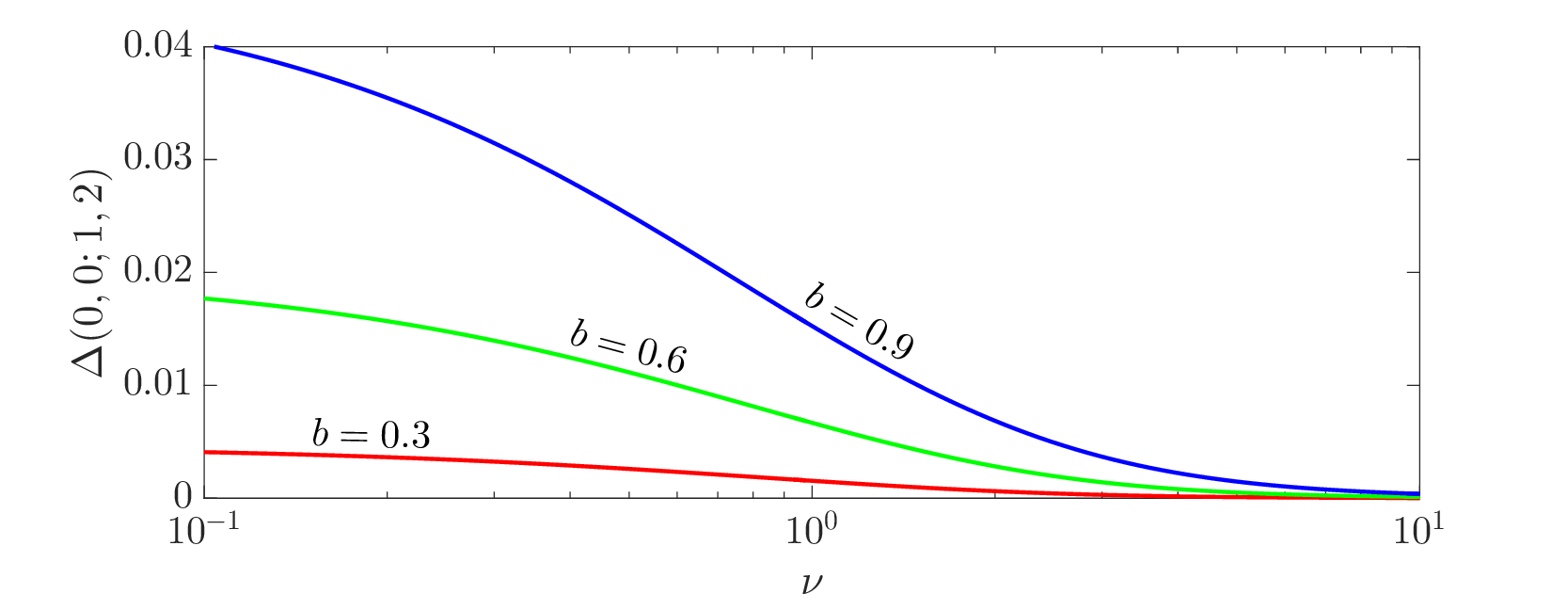}
    \end{flushright}\vspace{-6mm} 
  \caption{Plot of the difference between the LHS and RHS of \eref{eqn_chapman_kolmogorov}: $\Delta(x_0,x_2;t_1,t_2) := p(x_2;t_2|x_0;0) - \int dx_1\,p(x_2;t_2|x_1;t_1)\,p(x_1;t_1|x_0;0)$. One recognizes that $x(t)$ is not a Markov process. Only in the limiting cases $b\rightarrow 0$ and $\nu \rightarrow \infty$, does $x(t)$ become Markovian.}\label{fig:chapman_kolmogorov}
\end{figure}

We resort to demonstrating the violation of the Chapman-Kolmogorov equation for a special choice of $u=1/2$, $x_0=0$, $x_2=0$, $t_1=1$ and $t_2=2$. This is shown in \fref{fig:chapman_kolmogorov}. Therefore, $x(t)$ is in general not a Markov process. However, for the limiting cases of no coupling ($b\rightarrow0$) and time-scale separation ($\nu\rightarrow\infty$) the Chapman-Kolmogorov equation is fulfilled indicating that $x(t)$ becomes Markovian, as we would expect.

\end{document}